\documentclass[10pt,journal,compsoc]{IEEEtran}
\usepackage{graphicx}
\usepackage{color}
\usepackage{multirow}
\usepackage{amssymb}
\usepackage{paralist}
\usepackage{graphicx}
\usepackage{subcaption}
\usepackage{tcolorbox}
\usepackage{framed}
\usepackage[framemethod=default]{mdframed}
\usepackage{numprint}
\usepackage{url}
\usepackage[inline]{enumitem}
\usepackage{color}
\usepackage{colortbl}
\usepackage{xspace}
\usepackage{enumitem}
\usepackage{booktabs}

\usepackage{breakurl}
\usepackage[breaklinks]{hyperref}
\usepackage{url}

\usepackage{color}
\usepackage{rotating}

\newcommand{\quotes}[1]{\emph{``{#1}''}}

\newcommand{\miga}{{{MigA}}\xspace}

\newcommand{\new}[1]{{\color{black}{#1}}}
\newcommand{\todo}[1]{{\color{red}{#1}}}

\newcommand{\point}[1]{{\bf{#1:}}}

\usepackage{caption}
\usepackage{subcaption}

\usepackage{verbments}
\usepackage{xcolor}

\usepackage{listings}

\newcommand{\nremailreplies}{98\xspace}

\title{
Why did developers migrate Android applications from Java to Kotlin?
}

\date{}

\author{
Matias~Martinez,
 Bruno Gois Mateus\\
 Universit\'e Polytechnique Hauts-de-France, France and 
 LAMIH UMR CNRS 8201, France
 }

\newcommand{\rqwhymig}{{RQ 1: Why (for what reasons) have developers migrated their Android applications from Java to Kotlin?
}}

\newcommand{\rqwhynotmig}{{RQ 2: Why do developers not {fully} migrate their Android applications from Java to Kotlin?
}}

\newcommand{\rqadvise}{{RQ 3: What are the main takeaways from developers about migrating to Kotlin?}} 

\usepackage{subcaption}
\begin{document}

\maketitle

\thispagestyle{plain}
\pagestyle{plain}

\begin{abstract}

Currently, the majority of apps running on mobile devices are Android apps developed in Java.
However, developers can now write Android applications using a new programming language: Kotlin, which Google adopted in 2017 as an official programming language for developing Android apps.
Since then, Android developers have been able to:
\begin{inparaenum}[a)]
\item start writing Android applications from scratch using Kotlin, 
\item evolve their existing Android applications written in Java by adding Kotlin code (possible thanks to the interoperability between the two languages), or 
\item migrate their Android apps from Java to Kotlin.
\end{inparaenum}
This paper aims to study this last case.

We conducted a \new{qualitative study} to find out \emph{why} Android developers have migrated Java code to Kotlin and to bring together their experiences about the process, in order to identify the main difficulties they have faced.
To execute the \new{study}, we first identified commits from open-source Android  projects that have migrated Java code to Kotlin.
Then, we emailed the developers that wrote those migrations.
We thus obtained information from \nremailreplies{} developers who had migrated code from Java to Kotlin.
This paper presents the main reasons identified by the study for performing the migration.
We found that developers migrated Java code to Kotlin in order to access programming language features (e.g., extension functions, lambdas, smart casts) that are not available with Java for Android development, and to obtain safer code (i.e., avoid null-pointer exceptions).
We also identified research directions that the research community could focus on in order to help developers to improve the experience of migrating their Java applications to Kotlin. 

\end{abstract}

\renewcommand\IEEEkeywordsname{Keywords}

\section{Introduction}

Currently, Android from Google is the mobile platform used on most smartphones around the world~\cite{mobileshare}.
Traditionally, Android applications were developed using Java programming language.
However, in 2017, Google adopted Kotlin, a new programming language (v1.0 released in 2016), as an official language for developing Android applications~\cite{Shafirov2017LinkKotlinOfficialInAndroidJetBrains}.
Since then, Android developers have been able to develop Android apps using:
\begin{inparaenum}[a)]
\item Java, 
\item Kotlin, or 
\item both languages.
\end{inparaenum}

Kotlin is a programming language that combines object-oriented and functional features, some of them not present in Java or not available for Android development.\footnote{https://kotlinlang.org/docs/reference/comparison-to-java.html}
Kotlin is compiled to Java byte code, which means that 
\begin{inparaenum}[a)]
 \item an application written in Kotlin can be executed on the Java virtual machine (JVM), and
\item Kotlin is fully interoperable with Java, i.e., Kotlin code can invoke programs written in Java and vice versa.
\end{inparaenum}

The adoption of Kotlin as an official development language for building Android applications has resulted in three main scenarios.
Android developers can: 
\begin{inparaenum}[1)]
\item start writing an application from scratch in Kotlin,
\item evolve their Android apps, initially written in Java, by adding Kotlin code and maintaining the existing Java code,
\item totally migrate an application, initially written in Java, to Kotlin.
\end{inparaenum}
Recently, Coppola et al. presented the first characterization of migrations of Android apps from Java to Kotlin~\cite{Coppola:2019:CTK}.
They studied the evolution of such apps by using metrics based on the amount of Kotlin and Java code (LOC) and numbers of files. 
Their results show that the transition from Java to Kotlin was in general
\begin{inparaenum}[a)]
\item fast (rapid transition between languages), and
\item unidirectional (the ratio of Kotlin over total code was often increasing during their evolution).
\end{inparaenum}

In this paper, we go one step further on regarding the characterization of the migration of Android apps to Kotlin.
We conduct a \new{qualitative study} to study \emph{why} Android developers have migrated their applications from Java to Kotlin and bring together their experiences about the migration process.
This study is important for understanding the limitations of the traditional way of writing Android apps (i.e., using Java language) from the Android developers' perspective.
Moreover, we focus on developers who have begun migrating but have not finished at the moment of carrying out this study (i.e., their Android apps contain both Java and Kotlin code). 

To perform the \new{study}, we first executed code analysis to identify {commits} on open-source Android applications that have migrated Java code to Kotlin. 
From those commits, we found the developers that wrote them, i.e., that migrated code from Java to Kotlin.
Secondly, we contacted those developers via email to ask them why they migrated their Java code to Kotlin and their main reasons for doing so.

In total, we received responses from \nremailreplies{} developers that had migrated code on open-source Android apps published on app stores.
This study complements the one carried out by Oliveira et al. \cite{OliveiraSANER20p206}, where seven developers were interviewed to gain knowledge about the adoption of Kotlin in mobile development.
We also complemented the interviews with a study of the grey literature to detect problems related to the adoption of Kotlin that were not mentioned by developers.

A focus on migration from Java to Kotlin in Android development is crucial because we believe that, since the official adoption of Kotlin, Android development has entered a new era. 
We hypothesize that, given the increasing support that Google has given to Kotlin on Android, all Android development could eventually move from Java to Kotlin. 
For this reason, we consider that it is essential to understand {why} developers migrate Android apps from Java to Kotlin, in order to 
\begin{inparaenum}[a)]
\item uncover the difficulties developers have encountered and 
\item help and support them with documentation, techniques and tools for development and migration activities.
\end{inparaenum}
In this paper, we identify potential directions that the research community could focus on to help developers overcome such difficulties.

The contributions made by this paper are:
\begin{itemize}
\item A  detailed list of the main reasons that Android developers give for migrating to Kotlin.
This could encourage other Java developers to migrate to Kotlin.
\item A list of \new{experiences}, 
takeaways and recommendations from our \new{study}, that could be used by both Android developers (e.g., to decide whether to migrate to Kotlin or not) and by researchers (e.g., to propose solutions that overcome the current problems developers face).
\end{itemize}

The paper continues as follows.
Section \ref{sec:methodology} presents the methodology.
Section \ref{sec:evaluation} presents the responses of the research questions.
Section \ref{sec:tov} discusses the threats to validity.
Section \ref{sec:discussion} presents a discussion and future work. 
Section \ref{sec:rw} presents the related work.
Section \ref{sec:conclusion} concludes the paper.

\section{Background and motivation}
\label{sec:background}

In this paper, we focus on the migration of applications from Java to Kotlin.
This type of migration has some particularities, and differs from other types of migrations, e.g., of legacy systems.
A legacy system can be defined as ``{a system that significantly resists modification and evolution}''~\cite{Brodie1995Legacy}.
Bisbal et al.~\cite{Bisbal1997Migration} mention that legacy systems can cause problems, since they usually run on obsolete hardware and lack clean interfaces to interact with other systems.
The migration of Java to Kotlin has some key differences with legacy migration. For example, the underlying run-time environment (i.e., the Java virtual machine, ART and Dalvik machines for Android) does not need to be updated: Kotlin and Java are compiled to Java bytecode.
In addition, the communication between migrated and unmigrated components in legacy migrations (e.g., COBOL to web~\cite{Colosimo2009Evaluatinglegacy}) needs \emph{wrappers} (\cite{Bisbal1997Migration,Colosimo2009Evaluatinglegacy}) or \emph{gateways} (\cite{Brodie1995Legacy, Bisbal1997Migration}).
Conversely, 
the interoperability between Java and Kotlin means that wrappers and gateways are not necessary for migrating from Java to Kotlin.
Consequently, developers can, for instance, introduce Kotlin code into their Java apps without changing either the environment infrastructure or the Java code.
According to Oliveira et al. \cite{OliveiraSANER20p206}, developers seem to consider this interoperability one of the great benefits of adopting Kotlin, even though there are inconveniences and disadvantages to be tackled, such as some problems with language features and development tools. 

In any case, since it was declared an official language for Android development in 2017, Kotlin has been increasingly adopted (\cite{Coppola:2019:CTK, GoisMateus2019}) and some developers have decided to totally or partially migrate  their apps from Java to Kotlin.
The transition from Java to Kotlin was recently examined in two previous studies, one by Coppola et al.~\cite{Coppola:2019:CTK}, and another by Gois Mateus et al.~\cite{GoisMateus2019}.
Both studies report that once Kotlin is introduced into open-source Android applications that had initially been written in Java, most of them evolve by having more Kotlin code and less Java code.

Moreover, Coppola et al.~\cite{Coppola:2019:CTK} report that most projects that featured Kotlin in their latest release showed a rapid transition from Java to Kotlin during their lifespan. 

Beyond the recent progress on characterizing the development of Android apps that use Kotlin, we note that there are gaps in our understanding about migrations to Kotlin:
to our knowledge, no previous work has focused on capturing the motivations of migrating to Kotlin, or the experiences of developers that have already migrated code from Java to Kotlin in Android applications published in app stores.

This paper aims at filling those gaps.
In order to do this, the research questions that guide our study are the following:

{\rqwhymig}

{\rqwhynotmig}

{\rqadvise}

\section{Methodology}
\label{sec:methodology}

In order to respond to the research questions, we carry out a \new{qualitative study} with the main goal of collecting data from software developers about the migration of Java code to Kotlin in the context of Android development.

In this study, we choose to directly ask to the developers their motivations behind deciding to migrate from Java to Kotlin. 
The most suitable approach for obtaining this information is to employ qualitative data collection and analysis techniques, rather than other techniques such as experiments or quantitative surveys~\cite{saunders2009research}.

\new{
In Section~\ref{sec:methodology:overall} we present the overall research design.
Then, in Sections~\ref{sec:methodology:rq1},  \ref{sec:methodology:rq2} and \ref{sec:methodology:rq3} we describe the methods for responding to each research question.
}

\subsection{Overall Research Design}
\label{sec:methodology:overall}

Our research design was articulated as follows.

Firstly,
in order to sample the population that was able to participate in our study,
we performed code analysis to detect commits that migrated code from Java to Kotlin. 
Using these commits, we identified the developers who had written them.
This step is explained in Section \ref{sec:methodology:collecting_migrates}.

Secondly, we performed a qualitative study, consisting of contacting the previously identified developers with the goal of asking them about the migration process that they had carried out.
We contacted all the previously identified developers  by email, presenting them with a short semi-structured interview. 
We defined these interviews as semi-structured because the interviewees were free to answer the questions without pre-coding the answers, as usually done in a questionnaire~\cite{saunders2009research}.
The methods that we applied in executing the \new{study} are explained in Section~\ref{sec:methodology:survey}.

Thirdly, we carried out content analysis to highlight the developers' main motivations for opting to migrate from Java to Kotlin.
The methods we applied in this analysis are explained in Section~\ref{sec:methodology:recolections}.

Fourthly, 
we complemented the qualitative analysis described in Section~\ref{sec:methodology:survey} with an analysis of the grey literature about Kotlin migration.
This analysis aims to detect topics discussed and not discussed by developers during the mentioned interviews.
We present the methodology for collecting and analyzing the grey literature in Section~\ref{sec:methodology_greyliterature}.

\subsubsection{Identifying developers that have migrated code}
\label{sec:methodology:collecting_migrates}

\paragraph{Definition of migration commits}

In this work, a \emph{migration commit} is a commit that removes a piece of code written in Java and introduces code written in Kotlin.
For example, commit \texttt{3638be} from application \texttt{Chicago Commutes} removes a file named \texttt{GPSUtil.java}, written in Java and introduces a new one, \texttt{GPSUtil.kt}, written in Kotlin. 
Both files provide the same functionality: a method that returns the position given by a smartphone's GPS.\footnote{Migration commit: \url{https://github.com/carlphilipp/chicago-commutes/commit/3638be60c8bd144b968f044c0ded218e19697d69}}

\paragraph{File-level migration commit}
\label{sec:heuristics_detection_commits}

This paper focuses on one type of migration, which we call \emph{file-level migration}, that corresponds to commits that migrate code by removing one Java file and adding one new Kotlin file with the same name.

The heuristic for detecting such commits checks, for a given commit $C$, whether it removes a Java file named $F.java$ and adds a file named $F.kt$.

\paragraph{
\new{
Collecting migration commits using the \miga{} tool}}
\label{sec:methodoly_miga_implementation}

\new{To identify developers that have written migration commits from open-source repositories, we implement an open-source tool called \miga~\cite{migatool}.}

\new{
\miga is built on top of Coming \cite{coming2019}, a framework for studying the evolution of applications.
Coming provides functionality to, for instance, extract and analyze the source code changes introduced by each commit from a GIT repository.
In particular, \miga{} aims to detect commits that migrate code from Java to Kotlin.
}

\miga{} takes the location of a cloned GIT repository as its input.
It navigates each branch from the repository in chronological order, starting from the oldest one. 

\new{During the commit navigation, \miga{} analyzes each commit by calling a pipeline of commit analyzers. 
Commits belonging to several branches are analyzed only once.
For each commit, an analyzer from \miga{} inspects the source code files that are added, removed and modified by the commit. 
\new{More specifically, it applies the heuristic described in section \ref{sec:heuristics_detection_commits}: it stores a commit if it removes a Java file and also adds a new Kotlin file with the same name}.
}

Finally, it outputs a list of the migration commits previously filtered.
For each commit, \miga{} reports:
\begin{inparaenum}[a)]
\item commit ID (SHA-1),
\item developer's user name,
\item developer's full name,
\item developer's email,
\item branch(es) it belongs to, and
\item date.
\end{inparaenum}
That information allows us to contact developers that have migrated code.

\new{The architecture of \miga (inherited from Coming) allows users to add new commit analyzers via its extension mechanism. 
Using them, researchers could, for instance, encode  new heuristics in \miga{} for detecting migrations that are not covered by this paper.}

\subsubsection{
Data Collection
}
\label{sec:methodology:survey}

\new{
To address our research questions, we conduct a qualitative study~\cite{silverman2013doing}.
}
Our study is exploratory~\cite{experimentationSE2012} and applies short semi-structured interviews, \new{following a \emph{purposive sampling} strategy~\cite{patton1990qualitative} based on the data retrieved in Section \ref{sec:methodology:collecting_migrates}.}

First, from each migration commit previously retrieved, we glean the name and username of the developer that made the commit. 
Emails addresses are collected by querying GitHub API, using the username as input.

Then, we send a personalized email to each developer, that mentions that we have analyzed the code of her/his application (mentioning the app name), and that   we have detected that she/he wrote at least one commit that migrates code from Java to Kotlin.\footnote{Even if a developer has performed different migrations, we send only one email.}
We first ask developers a single question via email:
\quotes{Why did you migrate code from Java to Kotlin?}.
When we receive an answer from that developer,
we identify the main points discussed in the answer, using \new{content analysis}, and we ask new questions accordingly.
In addition to these questions, we ask other questions that we have predefined.
Examples of those are: 
\begin{inparaenum}[a)]
\item \quotes{Which Kotlin features do you like the most?}
\item \quotes{Did you use the auto-converter code tool provided by the IDE?},
\item \quotes{Which was the criterion (if any) to choose to migrate those classes?}.
\end{inparaenum}
The complete list of questions is available in our appendix~\cite{appendix}.

We choose a \new{qualitative study} format, composed of an initial question, and followed by a 
\new{semi-structured interview}, following the recommendations given by~\cite{experimentationSE2012}, 
\new{which indicates that giving the interviewee freedom to answer open-ended questions such as the ones proposed in our semi-structured interviews offers a number of advantages},
such as observing the answers and asking further questions according to what is observed.

Moreover, we choose to carry out a
\new{qualitative study}
because, as suggested by \cite{experimentationSE2012}, it can be used as a pre-study for a more thorough investigation, and may provide new possibilities that could be analyzed.
In Section \ref{sec:futurework}, we discuss new possibilities raised by our \new{qualitative study}.

\subsubsection{Collection of results and summary}
\label{sec:methodology:recolections}

To summarize the results from the study, 
we carried out content analysis~\cite{silverman2013doing}. 
We first carefully read through all the answers received, identifying their main motivations and assigning them a code. 
This process of coding allows us to summarize the answers from our interviewees inductively in order to discover new perspectives and insights.

\new{Our approach follows two ways of coding derived from grounded theory~\cite{glaser2017discovery}:}
Open coding and Axial coding.
Here we explain the steps.

\point{Open Coding}
The open-coding process consists of breaking down the content to be analyzed (e.g., responses to our emails) into different parts, each identified by a \emph{code}.
Then, we label them with words or short phrases, which we denominate \emph{codes}.
To carry out this step, we collect all codes discussed in each email thread (i.e., the original emails with answers and responses, if any).
For example,
from a statement such as:\quotes{I moved on from Java to Kotlin because I liked the features that are present in Kotlin and C++, but not in Java, [...] also because of extension functions} 
we extract the codes \texttt{Like features from C++}, and \texttt{Like new feature extension functions}.
More examples are listed in our appendix~\cite{appendix}.
At the same time, we also label the emails that contain valuable information for responding to the research questions, e.g., why developers migrated to Kotlin.

\point{Axial coding}
Once we have read all the emails and collected all the codes, we merge the codes that are equivalent or related. 
For example, the codes \texttt{Limitation of JVM/JDK} and \texttt{Java development stuck with old Java version} were extracted from two emails.
However, they refer to the same issue: migration due to the limitations of Java source code version (Java 6) when targeting all Android devices.
These are merged resulting in a new code: \texttt{Find Java a limitation from Android development}.
Once we obtain the final list of codes, we categorize them.
In total, we find 74 codes, which are available in our appendix~\cite{appendix}.

\point{{Classification of emails}} 
\new{Once we obtain the final list of codes, we look through all the emails again to assign each with a code.}
To do this, we create a spreadsheet where each column is a code $C$ and each  row corresponds to a developer $D$ that we have interviewed. 
Then we read each email $E$ from developer $D$: for each code $C$, we mark the cell $D,C$ if the code is present in any of the emails from developer $D$. 
Otherwise, we leave that cell empty.

\subsubsection{Analysis of the grey literature}
\label{sec:methodology_greyliterature}

\new{
We complement the qualitative study with a \new{secondary analysis~\cite{saunders2009research} of the grey literature}.
The main goal of this analysis is to identify which of the reasons for migration mentioned by developers were also mentioned by the grey literature.
}
In this paper, we focus on tutorials and blog entries.

To collect them, we execute two queries on Google: 
\begin{inparaenum}[\it 1)]
\item \texttt{Android Java Kotlin migration}, and
\item \texttt{Android Java vs Kotlin}.
\end{inparaenum}
Then, we inspect the top-25 sites returned, and apply the methods explained in Section~\ref{sec:methodology:recolections} to extract codes from them.
Our appendix includes the links to all sites analyzed in the grey literature.

Please note that the study of the grey literature does not aim to report the codes that are most discussed, but to detect those that the developers mentioned. 
To respond to our research questions, we carry out a secondary analysis~\cite{saunders2009research} based on an additional source (the grey literature), rather than a triangular study (which involves using two or more methods of data collection ~\cite{cohen2002research}).

To complement the study, in the discussion section (\ref{sec:differencesgreyliterature}), we present codes from the grey-literature that have not been mentioned by developers.

\subsection{\new{Method} for responding to RQ 1: Why have developers migrated their Android apps to Kotlin?}
\label{sec:methodology:rq1}

To respond to  RQ 1, we inspect each code $C_i$ (determined by using the method explained in Section~\ref{sec:methodology:recolections}), and decide whether the code could be a reason for migrating or not.
If it is, we read all the emails associated with code $C_i$ and validate if the developer who wrote each email mentions $C_i$ as a reason. 
It should be noted that it might be the case that some codes are related to one another. 
For example, one code states that developers like the Extension Functions feature, and another code does the same with the Data Classes feature.
In those cases, we report a single reason that covers both codes.
\new{Finally, we report the 10 most frequent reasons that we find.}

\subsection{\new{Method} for responding to RQ 2: Why not fully migrated?}
\label{sec:methodology:rq2}

To identify why some developers have not yet finished a migration, we first identify the applications that have been \emph{partially} migrated: ones whose latest versions contain both Kotlin and Java.
In order to do so, we analyze the amount of code written in Java and in Kotlin from the first and last commit in each app.
If the application has Java code in the initial version but none in the last one, we consider it \emph{fully} migrated.
Conversely, if the application has both Java and Kotlin code in the latest one, we consider it \emph{partially} migrated.

Then, we apply the methods explained in Section~\ref{sec:methodology:collecting_migrates} to the set of partially migrated apps\footnote{At the moment the study was done, the partially migrated apps analyzed still had Java code.}, to retrieve developers that have migrated code.
Finally, we ask them about their reasons for not having finished the migration process, and why they still have Java code in their codebases.

\subsection{\new{Method} for responding RQ3: Migration takeaways}
\label{sec:methodology:rq3}

To answer RQ 3, we follow the same methods as for responding to RQ 1 (Section \ref{sec:methodology:rq1}), but include an additional step.
When we manually analyze each email, we also capture the codes that are related to the experiences of the developers during the migration.
For example, from the statement: \quotes{Auto-converter does its job only half, still need to go through and fix some things}, we mark that the developer wrote a takeaway related to the code: \emph{Had issues when converting using auto-converter}.

\subsection{The data under study}

In this section, we first present the applications that we analyze to respond to these research questions.
Then, we report the amount of migration commits we find in those apps. 

\subsubsection{The dataset of Android applications}
\label{sec:dataset}

We define two main construction criteria in our dataset of applications.
First, the version history of the applications to study must be tracked on the Git system and publicly available in development platforms such as GitHub.
These criteria allow us to apply the heuristic discussed in Section \ref{sec:heuristics_detection_commits}.
Second, to avoid analyzing ``toy" projects that developers might upload to such platforms, 
the apps must be published on an app market, i.e., official ones such as Google Play or unofficial ones such as F-Droid.

To collect the largest number of open-source Android applications, 
we focus on two datasets: 
\begin{inparaenum}[\it 1) ]
\item F-Droid\footnote{F-Droid: \url{https://f-droid.org}}, an app market of open-source apps,
and 
\item Android Time Machine~\cite{Geiger2018:data}, a dataset of apps published on Github.
\end{inparaenum}

In total, we collect 2,167 open-source Android applications that fulfill the criteria mentioned above. 
Then, we use the heuristics for filtering Kotlin applications presented in~\cite{GoisMateus2019}.
At the moment the study was performed, the partially migrated apps analyzed contained Java code.
We find 374 out of 2,167 that have at least one line of code written in Kotlin. 
Those apps have a median of:
\begin{inparaenum}[a)]
\item 331 commits,
\item 139 files,
and
\item 7951 lines of code.
\end{inparaenum}
More statistics are available in our appendix~\cite{appendix}.

To respond to RQ 2 (why apps are not fully migrated), we first detect the applications that have not been fully migrated.
We find 214 out of 374 apps initially written in Java that have been partially migrated (i.e., that have both Kotlin and Java code in the latest commit).

\subsubsection{Finding Migration Commits and Developers}

We execute \miga{} on the 374 apps that have at least one commit with at least one line of code written in Kotlin.
We mine commits from all branches, following the suggestion from~\cite{Kovalenko2018Branches}.
In total, we find 3520 migration commits written by 362 different developers.
95.7\% of these commits (3368/3520) belong to the active branch (usually named \emph{master} or \emph{development}). 
The remaining 152 commits belong to other unmerged branches. 
Finally, from those migration commits, we retrieve the usernames, full names, and emails of the developers.
In total, we find 362 distinct developers, who will form the target of our study.

\section{Results from the qualitative study}
\label{sec:evaluation}

\begin{table*}[h!]
  \begin{center}
  \small
    \caption{The most frequent codes mined from the survey with developers. Annotations RS, RNM, and RT refer to codes used for responding to research questions 1, 2 and 3, respectively. 
    \new{Those numbered are detailed in Section~\ref{sec:evaluation}.}
   \new{The full table is available in our appendix~\cite{appendix}}.}
    \label{tab:concepts}
    \begin{tabular}{llr} 
    \toprule
      Category	& Code			& \#Dev	\\
      \midrule
    Support	
	&	Like Google support 	\new{(RS: 2)}	&	32  	\\
	&	Like JetBrains support	\new{(RS)}	&	8	\\
	&	Highlight Importance of IDE for Kotlin		&	6	\\
	&	Adopted due to coolness factor	\new{(RS)}	&	4	\\
	\hline
Advantage of Kotlin 
	&	Reduce code	\new{(RT: 1)}	&	34	\\
code with respect  	&	Easier to read/clear/write/syntactic sugar/easier to maintain	\new{(RT: 1)}	&	23	\\
 to development &	Kotlin has clear syntax		&	17	\\
and maintainability		&	Less Redundant/less verbose 		&	12	\\
	&	Little-to-no boilerplate code		&	9	\\
	&	Kotlin is a modern language/has features from modern languages	\new{(RS)}	&	5	\\
	&	Code more dense/heavy		&	4	\\
	&	Efficient language		&	2	\\
	\hline
Design
	&	Ease of implementing MVI (Model-View-Intent) architecture	\new{(RS:3)}	&	6	\\
	&	Ease of implementing MVVM (Model-View-ViewModel) architecture \new{(RS)}		&	1	\\
	&	Availability of Built-in design patterns (e.g., with Extension functions)		&	1	\\
	\hline
Relation with other 	
	&	Don't want to use Java anymore	\new{(RS: 7)}	&	13	\\
languages	&	Find Java a limitation for Android development	\new{(RS:8)}	&	10	\\
	&	Like Scala	\new{(RS: 6)}	&	5	\\
	&	Like Kotlin Features from C++		\new{(RS: 10)}	&	4	\\
	&	Kotlin has more extensions than Java native package		&	3	\\
	&	Don't like Java	 	\new{(RS)}	&	2	\\
	&	Like Koltin Features from Haskell		\new{(RS)}	&	2	\\
	&	More similarity with Java w.r.t other JVM languages		\new{(RS)}	&	2	\\
	&	Not a fan of Kotlin		&	2	\\
	&	Like  Python		&	1	\\
	&	Similar to Swift	\new{(RS)}	&	1	\\
	\hline
Language	
Features	&	Like Null safety/Nullability support		\new{(RS:1)}	&	42	\\
	&	Like Extension functions		\new{(RS: 3)}	&	23	\\
	&	Like Coroutine		\new{(RS:3, RT: 2)}	&	21	\\
	&	Like Data class (POJO)		\new{(RS: 3)}	&	16	\\
	&	Highlight Interoperability with Java	\new{(RS: 5)}	&	15	\\
	&	Like lambdas		\new{(RS: 6)}	&	14	\\
	&	Like Functional programming available in Kotlin	\new{(RS: 6)}	&	11	\\
	&	Use Coroutines instead of ReactiveX	\new{(RS)}	&	8	\\
	&	More features than Java 	\new{(RS)}	&	7	\\
	&	Like Function as parameters/Higher-Order functions		&	6	\\
	&	Like Immutability		&	5	\\
	&	Highlight Kotlin standard/built-in library		&	5	\\
	&	Like no getters/setters		&	4	\\
	&	Like streams		&	4	\\
	&	Like  built-in API for manipulating collections		&	4	\\
	&	Like default parameters		&	4	\\
	&	Find Kotlin safer than Java		&	3	\\
\hline
Productivity with Kotlin	
	&	Increase productivity		&	5	\\
	&	Faster development		&	4	\\
	&	Faster delivery		&	3	\\
	\hline
Running platforms
	&	Like Kotlin to write server side apps\footnote{We don't discuss this code in this paper as we focus on Android development.}	\new{(RS)}	&	7	\\
	&	Interested in multiplatform with Kotlin	\new{(RS:9)}	&	6	\\
	&	Kotlin Android Extensions plugin (Android KTX)		\new{(RS)}	&	2	\\
	\hline
Auto-converter
	&	Use the auto-converter alongside the migration process		&	36	\\
	&	Had issues when converting using \new{the} auto-converte\new{r} \new{(RT: 2)}		&	24	\\
	&	Use \new{the a}uto-converter first, then manual conversion		&	2	\\
	\hline
Migration process
	&	App migrated to learn Kotlin	\new{(RS: 4)}	&	19	\\
	&	Migrating using "Boy scout rule" (RNM: 1)		&	14	\\
	&	Only new features would be implemented in Kotlin	(RNM: 2)	&	10	\\
	&	Migrated in one step (i.e., one single commit)		&	9	\\
	&	Migrated only by hand (only manual coding)		&	6	\\
	&	No time to finish	(RNM: 3)	&	5	\\
	&	Only new files in Kotlin/Old code kept in Java	\new{(RNM)}	&	3	\\
	&	Started migration with data class (POJO)		&	2	\\
\hline
Performance 
	&	Affirm no overhead in performance 		&	2	\\
	&	Affirm no overhead APK size		&	1	\\
	\bottomrule
    \end{tabular}
  \end{center}
\end{table*}

We received responses from \nremailreplies{} developers, $\sim$27\%  of the total contacted (98/362), with all having developed distinct applications.

Table~\ref{tab:concepts} shows an extract of the summary of the codes obtained using the methods described in Section \ref{sec:methodology:rq1}.
The complete table is presented in our appendix~\cite{appendix}.
We now respond to the research questions based on those results. 

\subsection{\rqwhymig}

We present the 10 most \new{frequent} reasons for migrating Java code to Kotlin that we were able to identify from the interviews.
\new{We recall that a qualitative study (like ours) is not statistically representative of the whole population under study. Consequently, what we find are the 10 most relevant reasons among the interviewees}.

Each reason, $i$, is related to a code labeled as `RS:~$i$' in Table~\ref{tab:concepts}.
In this section, reasons  are sorted according to the number of developers that mention each reason.
We also group the codes according to different categories: e.g., Support (i.e., from Google and/or JetBrains), Design, and Language features.

\subsubsection{Reason: to avoid errors by using safer code \new{(RS: 1)}}
\label{sec:safercode}

One of the biggest problems in Java is the way it handles \emph{nulls}.
Incorrect manipulations of them lead to \emph{java.lang.NullPointerException} (NPE). 
As reported by Coelho et al. \cite{Coelho2015ExceptionsAndroid}, 
\emph{java.lang.NullPointerException} was the most reported root cause (27.71\%) found in issues reported in over Android projects.
Moreover, they found that 51.96\% of those projects reported at least one exception stack trace on which the NPE was the root cause.

Kotlin eliminates the possibility of empty pointers from a compilation perspective: potential NPEs are detected at compile time instead of crashing apps at runtime.

Forty-two developers stated that they decided to migrate to Kotlin to obtain safer code (i.e., \emph{null safety}).
For example, one developer wrote:
\quotes{I researched a bit on the language and the focus on null safety and immutability sold me on trying to avoid some past bugs}.
Another says: \quotes{One of the reasons that motivated me to migrate to Kotlin is Null Safety: reduces errors and I don't have to think if an object may become null}.

This reason is related to the \emph{maintainability of applications} and is discussed by articles from the grey literature, e.g.~\cite{Butterworth2020LinkMigration, Caneco2017LinkMigrating}.

\subsubsection{Reason: to follow Google~\new{(RS: 2)}}

Thirty-two developers told us that one of the main reasons for migrating to Kotlin was the fact that Google had adopted it as an official Android programming language.\footnote{Kotlin official language for Android: \url{https://developer.android.com/kotlin}}
This adoption means that Google: 
\begin{inparaenum}[a)]
\item expands documentation, resources and support for Kotlin development,
\item enriches the IDE Android Studio to support Kotlin,
\item provides \emph{Android KTX}, a set of Kotlin extensions for the Android platform.
\end{inparaenum}
One developer told us:
\quotes{I liked Kotlin but Kotlin was an unofficial language for Android development in Dec 2016.
But after one year, Google announced Kotlin was first-class language for Android development.
So no more worry about Kotlin being banned and I migrated Java to Kotlin}. 
Another developer told us:
\quotes{Kotlin being announced as the primary language by Google, we wanted to keep the product up to date with the latest innovations available in the market}.

Google's adoption of it boosted the popularity of Kotlin~\cite{OliveiraSANER20p206}.
For example, one developer told us: 
\quotes{I got interested in Kotlin after I saw some Kotlin snippets in the Android API documentation so I looked into it a bit more and I liked some of the codes I saw}.

Eight developers also remarked on the role of JetBrains, the company that created both the Kotlin language and the IDE IntelliJ IDEA. 
One of them told us:
\quotes{Another reason that influenced me into deciding to migrate to Kotlin was the support Kotlin has, being backed up by JetBrains and Google [...]. Since Google is pretty much the Android authority, it's wise to follow the best practices they recommend}.

Also, developers underlined the importance of the IDE Android Studio (which is based on the JetBrains' IDE IntelliJ IDEA).
One told us: \quotes{The IDE support, debug information, and bytecode viewer is a killer for senior Android devs to play with}.
There is one prominent functionality that it provides: a tool for automatically converting a Java file to Kotlin \quotes{with the click of a button}. 
One developer told us:
\quotes{Thanks to JetBrains, I migrated it in several hours, and it worked on the first launch, which was very promising}.

This reason is related to the \emph{Android development support} and discussed by articles in the grey literature such as \cite{daga2018LinkJavaVsKotlin, Sommerhoff2015LinkKotlin,Heath20LinkSwitch}.

\subsubsection{Reason: waiting to use a modern programming language \new{(RS: 3)}}
\label{sec:modernlanguage}

Most developers who answered our questions found Kotlin to be a \emph{modern} programming language that provides several built-in features that are not available natively in Java.
Some of them are:
\begin{inparaenum}[a)]
\item Extension functions (23 developers highlighted this),
\item Coroutines (21)
\item Data classes (16), 
\item others features (Smart casts, Type inference, Control flow).
\end{inparaenum}
For instance, the \emph{Extension function} feature provides the ability to extend a class with new functionality without inheriting from the class.\footnote{Kotlin Extensions: \url{https://kotlinlang.org/docs/reference/extensions.html}}
One developer wrote this about it: 
\quotes{apart from providing a clean way to refactor the code, extensions allow an alternative to Abstraction/Inheritance in order to achieve the Open/Closed Principle, which is invaluable}.

Six developers remarked on the \emph{built-in design patterns} provided by Kotlin, which enforce some of the best practices of Java by design.
For example, the \emph{Singleton} pattern using the keyword `Object' and the \emph{Decorator} pattern using the keyword `by'.

This reason is related to the \emph{development} of applications.
The grey literature we inspected also highlighted those features, for example, extension functions were mentioned by~\cite{daga2018LinkJavaVsKotlin,
Sommerhoff2020LinkJavaVsKotlin,
Rishabh2017LinkKotlinAndroid,
Thornsby2019LinkJavaVsKotlinKeyDiff,
Sommerhoff2015LinkKotlin}, 
and coroutines by~\cite{daga2018LinkJavaVsKotlin,Thornsby2019LinkJavaVsKotlinKeyDiff,Hiral2018}.

\subsubsection{Reason: for learning purposes~\new{(RS: 4)}}

Sixteen developers mentioned to us that they started developing their apps with for learning purposes: 
they generally try and test new technologies such as Kotlin while they develop an app.
(Note that, beyond that this purpose, as our inclusion criteria indicate (Section \ref{sec:dataset}), all analyzed apps were published on apps stores such as Google Play).

Those developers migrated Java code to Kotlin while they were learning to program in Kotlin.
One of them told us:
\quotes{I migrated my mobile application from Java to Kotlin mainly for learning purposes. Even though this project is in production, it is a great playground to stick with the last Android technologies and experiment new tools}. 

This reason is related to developers' \emph{training education and skills} and is discussed by articles in the grey literature, e.g., by \cite{Sommerhoff2020LinkJavaVsKotlin}.

\subsubsection{Reason: to use a new language that is 100\% interoperable with Java \new{(RS: 5)}}
Fifteen developers mentioned the advantage of full interoperability between Kotlin and Java.
This interoperability allows developers to \emph{mix} Java and Kotlin code. 
Thus, the migration can be performed gradually, i.e., a commit migrates some classes which can interact with unmigrated classes.
As one developer said: \quotes{We could implement new functionality in Kotlin while leaving the existing Java classes intact. Otherwise, the port to Kotlin would have been infeasible}.
This is aligned with the finding from Oliveira et al. \cite{OliveiraSANER20p206}: their interviews with seven developers present interoperability as a great benefit of adopting Kotlin.
Interoperability is a key factor for conducting  low-risk migrations by \emph{gradually} migrating the code. 
For instance, the Android app of Duolingo\footnote{Duolingo: \url{https://www.duolingo.com/}}, a language education platform, was fully migrated from Java to Kotlin in two years~\cite{duolingo_website}. 
This was possible due to the interoperability of the languages: during that period, Duolingo apps contained both Kotlin and Java code. 
As Duolingo's developers report, this gradual migration allowed them to apply strict testing, code review and code style of each part of the migrated application.

This reason is related to the \emph{development} and \emph{maintainability} of Android applications.
The grey literature also highlights interoperability as one of the most important features, e.g., \cite{daga2018LinkJavaVsKotlin, Rishabh2017LinkKotlinAndroid, Caneco2017LinkMigrating}.
Moreover, Sommerhoff \cite{Sommerhoff2020LinkJavaVsKotlin} states that, thanks to interoperability, the migration of large projects can be performed gradually.

\subsubsection{Reason: to use a functional programming language for Android development \new{(RS: 6)}}

Eleven developers migrated to Kotlin because it is a \emph{functional-oriented} programming language, and it provides several \emph{functional} features  that are not available for Android development using Java 6 (e.g., lambda is available in Java 8+).
The \emph{Higher-Order functions} feature (i.e., the possibility of passing a function as an argument) was highlighted by fourteen developers.
Five developers told us they were mainly Scala developers.
Kotlin gives those developers the possibility to write Android applications using a functional paradigm, just as they do with Scala.

This reason is related to the \emph{development} and \emph{maintainability} of applications. Several articles from the grey literature highlight functional features as one of the main advantages of adopting Kotlin, e.g., \cite{daga2018LinkJavaVsKotlin,Sommerhoff2020LinkJavaVsKotlin, Sommerhoff2015LinkKotlin}.

\subsubsection{Reason: to avoid Java language \new{(RS: 7)}}
Some developers migrated their tools because, as they confessed, they do not like the Java language.
Thirteen developers mention that, thanks to the possibility of coding Android apps with Kotlin, they no longer use Java.
For example, one wrote: 
\quotes{I don't and haven't ever really liked Java. I only wrote it in Java because that's what Android dictated. When Kotlin came around, it looked like a nicer language, and was fully compatible}.

Since Google officially adopted Kotlin, developers who do not like or do not use Java can now develop Android apps in another programming language.
One of them wrote:
\quotes{I migrated from Java to Kotlin because I don't really know how to work with Java. The app was written in Java because there was no officially supported alternatives for Android development}.

This reason is related to the \emph{development} of applications.

\subsubsection{Reason: to avoid Android platform fragmentation and the limitation of Java versions used for Android development~\new{(RS:~8)}}

One of the main obstacles that Android developers face is the fragmentation of the Android platform.
To target all platforms, Android developers who use Java are obliged to use Java 6, which does not include coding features such as \emph{lambdas and extension functions}. 
We call them \emph{modern features} as they are typically available on modern programming languages such as Scala, Golang, Rust, Swift, etc. 
To use Java 8 and its modern code features (including collection API, streams and lambdas), their apps must target one of the latest Android OS versions (API level 24+).\footnote{Java 8 support: \url{https://developer.android.com/studio/write/java8-support}}
This implies that Android devices with older OS versions cannot run applications written using Java 8. 

Ten developers noted their annoyance about programming on Android using Java 6.
One of them said:
\quotes{I was sick of using Java. As a professional Android developer there is no reason to use Java, especially the incredibly limited version of Java you get on Android.} 

The majority of the developers that answered us stated that migrating to Kotlin was a way to ``hack'' this problem: 
Kotlin has standard libraries that provide features available on Java 8+ (such as lambdas and streams).
Gois and Martinez \cite{Gois2020Features} describe these Kotlin code features that are not available in Java for Android development.
Moreover, Kotlin has as advantage that its source code is compiled to Java 6 bytecode. 
Thus, apps written in Kotlin can be executed on any Android device. 
This simplifies the development task.
For example, one developer wrote: 
\quotes{I forked a library into my project and it was Java 8 only because it used streams, but with Kotlin I got that working in Java 6/7 which is how most android works}. 

This reason is related to the \emph{development} and \emph{deployment} of Android applications and is also discussed in the grey literature, for example, by \cite{daga2018LinkJavaVsKotlin, Kust2017LinkKotlinFeatures}.

\subsubsection{Reason: to achieve multi-platform development \new{(RS:~9)}}
\label{sec:reason:multiplatform}
It is worth mentioning that, even though no developers mentioned it as their main reason for migrating, 
seven of them pointed out that the migration to Kotlin could allow them to achieve multi-(cross-) platform development.\footnote{Kotlin multiplatform: \url{https://kotlinlang.org/docs/reference/multiplatform.html}}
Currently, mobile developers can write the business logic of a mobile application using Kotlin and share it in their Android and iOS projects, allowing them to reduce development time and effort by reusing business code.\footnote{Mobile cross-platform: \url{https://www.jetbrains.com/lp/mobilecrossplatform/}}

This reason is related to the \emph{portability} of Android applications.

\subsubsection{Reason: to use features provided by other programming languages \new{(RS: 10)}}

Some developers told us that they master or prefer other non-Java Virtual Machine (JVM) programming languages such as C, C++, Python or Haskell.
As Android developers have been ``{forced}'' to program in Java, the introduction of Kotlin for Android development has been an opportunity for those developers to use some features also available in their favorite programming languages.
\emph{Named and default parameters} in Python and \emph{Operator overload} in C++ are features present in Kotlin that developers mentioned in their answers.

This reason is related to the \emph{development} and \emph{maintainability} of applications.
The grey literature also highlights these features of Kotlin, e.g., named and default parameters~\cite{Sommerhoff2020LinkJavaVsKotlin,Gour2020,
Sommerhoff2015LinkKotlin}, and
operator overload~\cite{Kust2017LinkKotlinFeatures,
Hiral2018}.

\subsection{\rqwhynotmig}
\label{sec:whynotmigrate}
Thanks to the interoperability between Kotlin and Java, developers do not need to migrate their applications completely.
We asked developers of not fully migrated apps why they had not yet finished the migration.
We set out the most important reasons identified. 
Each is related to a code labeled as `RNM:~$i$' from Table~\ref{tab:concepts}.

\subsubsection{The ``Boy Scout" rule \new{(RNM: 1)}} 
Fourteen developers migrated Java code to Kotlin by following the \emph{``Boy Scout Rule''}.\footnote{The name `Boy Scout Rule' was coined by a developer during the interview. 
}
This rule states that the code is migrated if:
\begin{inparaenum}[a) ]
\item it is necessary to change a Java file, and 
\item migrating it to Kotlin is simple (i.e., it takes little time).
\end{inparaenum}

Unlike from those developers that completely migrated their code, some developers of partially migrated apps only migrated code when there was a particular reason for doing so.
For instance, some of these developers told us that they migrated while they were refactoring code.
As one wrote: \quotes{Occasionally parts of the application that need refactoring are identified. When refactored, they are migrated to Kotlin}.
Another refactored while converting: 
\quotes{I converted whenever I had to touch the file for one reason or another anyway, usually combined with a refactoring}.
This way of migrating causes the amount and the proportion of \new{Kotlin} code to grow along as the apps evolve.

However, some developers remarked that they do not always follow that rule.
For instance, one told us: 
\quotes{Old classes were migrated to Kotlin when they had changes. Some classes are still in Java because they implement hard logic}.

The grey literature also mentions the style of migration that follows the ``Boy Scout" rule. 
For instance, Abdelaziz \cite{AbdelAziz2020MigEnterprise} suggests: 
\quotes{when a Java class raises a bug and needs to be changed, [...] then you can also convert it on the spot.}

\subsubsection{New functionality is written in Kotlin, old functionality remains in Java \new{(RNM: 2)}}
Some developers told us that they decided to write only new functionality in Kotlin without migrating the \emph{old} Java code. 
In other words, components written in Java evolved without any migration, unlike the ``Boy Scout'' rule.
One told us: 
\quotes{I don't think I have yet changed any code just for the sake of rewriting}.
Upon encountering a bug in a Java file, these developers keep the Java code.
One mentioned: \quotes{I decided that every new feature would be implemented in Kotlin. I would add Java code only to fix legacy code}.
Another developer suggested:
\quotes{not touching the old code unnecessarily. Adhere to the principle of `work, do not touch'. And only when expanding the functionality of the old Java code, translate it into Kotlin}.
Similarly, one told us: \quotes{If it was working in Java, it'd continue to work}.

This strategy was also discussed in the grey literature. For example, a post from a development company~\cite{strumenta2020} recommends, according to their experience, \quotes{not to convert Java to Kotlin when your existing Java code base is large and in maintenance mode [...] you are not going to save much by putting old code in a new form}.
Moreover, this strategy was also observed from a survey by Khadka et al. on legacy system modernization \cite{Khadka2014LegacyModernization}.
There, some practitioners and respondents to our interviews indicated that \quotes{if the legacy systems are working well, then legacy system modernization projects are unlikely to be initiated}.

\subsubsection{Time factor \new{(RNM: 3)}} 
Some developers told us that their applications have still not been migrated because
they simply have not had time to finish the migration.
This aligns with the finding from the survey by Khadka et al. on legacy system modernization~\cite{Khadka2014LegacyModernization}. 
They reveal that, according to the practitioners they interviewed, \quotes{finishing any legacy system modernization on time is the biggest challenge}.
Others told us they are not in a hurry to complete the migration.
Finally, some developers are waiting for the release of particular features from the Kotlin/Android platform.
For example, one developer told us: 
\quotes{I am not in a hurry to get to `100\% Kotlin', at least not until the future of `Kotlin Native' (for a possible iOS port) is clear}.

\subsection{\rqadvise}
\label{sec:advidses}

We list three takeaways reported by developers about their experience \emph{after} migrating to Kotlin.
They are related to codes labelled as `RT:~i' in Table~\ref{tab:concepts}.

\subsubsection{Adoption of Kotlin produced less, and clearer, code~\new{(RT 1)}} 
\label{sec:advise:lesscode}

Thirty-four developers remarked that Kotlin allowed them to write less code than in Java.
Seventeen mentioned that Kotlin has better and more precise syntax than Java, producing less verbose and less redundant code.
As one told us: \quotes{Kotlin is a very concise and expressive language} and 
\quotes{Kotlin strikes a good balance between being concise vs cryptic}.
Articles from the grey literature also remark on the reduction in lines of code when they compare an application written in Java with its equivalent written in Kotlin.
However, the reduction of lines of source code written differs across the articles.
For instance, Trehan \cite{Trehan2019Migration} and Wilson \cite{Wilson2019Metrics} both report a reduction of around 7\%, Caneco \cite{Caneco2017LinkMigrating} reports a 23\% reduction, and the reduction observed in the experiment from Uber~\cite{Fernandes2019UberPerformance} is 40\%.
This divergence is a call to do more extensive research on the measurement of code reduction due to adopting Kotlin.

Nine developers remarked that Kotlin allows \emph{boilerplate code} to be reduced, i.e., code that has to be included in many places with little or no alteration.
One of the features which aids that reduction is \emph{Data classes}: a model class can be declared in one line.
Sixteen developers included \emph{Data classes} in their favorite Kotlin features.

Furthermore, twenty-three developers remarked that Kotlin code is easier to read, write and maintain than Java code.
Twelve developers mentioned that Kotlin has a simpler and clearer syntax than Java.
(We recall that \emph{all} developers we contacted wrote their applications using Java \emph{before} migrating -totally or partially- to Kotlin).

Nevertheless, among the \nremailreplies{} developers who replied to us, two developers mentioned that, when they started programming in Kotlin, its compact syntax affected code readability.
One told us: \quotes{Initially, it felt awkward, but after a month of practice with the language, it was already obvious that the syntax was much improved over Java}.
These findings align with those of Oliveira et al. \cite{OliveiraSANER20p206}, which found that the overuse of lambdas and closures can decrease code readability.

{\bf Implication:} 
The developers agreed that
Kotlin allowed them write more concise, less redundant and more precise code. 
Kotlin helps developers to write and maintain their Android applications.
However, Kotlin's concise syntax might cause difficulties for novice developers.

\subsubsection{Careful use of auto-converter provided by the IDE~\new{(RT: 2)}}
\label{sec:autoconverter}
In total, thirty-six developers told us that they had used the code converter provided by the IDE, which converts a single Java file to Kotlin.
Some of them indicated that they had started using it while they were learning to code in Kotlin, but then, after gaining confidence, they continued the migration without using the converter.
All developers who used the conversion tool told us that they modified the code after conversion.
They agreed that the converter helped them to execute the migration and, in general, that the changes that needed made to the converted code were simple.
Unanimously, the  main reasons  mentioned for making such changes after the conversion are:
\begin{inparaenum}[\it 1)]
\item to transform nullable variables into non-nullable, and
\item to make the converted code more idiomatic.
\end{inparaenum}

Related to the nullability of variables, developers mentioned that they had to modify the converted code to better support nullability. For example, one told us: \quotes{It was also marking variables and properties as nullable too often, I was able to make them non-nullable after the conversion}. 
We recall that Kotlin, by default, does not allow a variable to hold a null, and null check is done at compilation time.
The developer, however, can declare nullable variables by using the operator `?' e.g., \texttt{val number: Int? = null}.
Then, using the variable, a developer has two main options: 
\begin{inparaenum}[\it 1)]
\item to use the safe operator~`?' i.e., \texttt{number?.toString()}: the method \texttt{toString()} is not invoked if the variable is null;
\item to use the unsafe operator~$!!$ \texttt{number!!.toString()}: the method is always invoked, and if \texttt{number} is null, it throws an exception. 
Thus, developers should do something that is typically done in Java: add a guard (an if statement) which checks for a null reference before accessing that variable.
\end{inparaenum}
The auto-converter from the IDE outputs the unsafe operator $!!$ in every variable access of potentially nullable variables.
Consequently, to make that transformed code more Kotlin-idiomatic (i.e., Null safety, one of the main features of Kotlin), developers would need to remove those $!!$ operators.
Articles from the grey literature, e.g.,~\cite{Vavra2017,Baxter2017} present different strategies to remove such $!!$ operators.

{\bf Implication:} 
The auto-converter tool provided by the IDE allows Android developers to obtain an \emph{initial} version of their applications in Kotlin.
However, it is necessary to know Kotlin well in order to modify the  code generated to make it more idiomatic.

\subsubsection{Simplification of asynchronous tasks using Coroutines~\new{(RT: 3)}}

Asynchronous or non-blocking programming is essential on Android because it allows better user experience and improves application performance. 
It is used to perform network calls, execute background jobs and tasks, access the local database, and run computationally intensive calculations. 
There are several manners of implementing asynchronous tasks in Android development with Java: AsyncTasks, plain old threads, Android’s main looper, Android loaders, etc.
To simplify the development of asynchronous tasks, some Android developers use external libraries (i.e., non-native concurrency API)
such as RxJava.\footnote{RxJava: \url{https://github.com/ReactiveX/RxJava}} (The survey carried out by Verdecchia et al.~\cite{Verdecchia2019Guidelines} shows that RxJava was the Java library that was most mentioned by the Android practitioners they interviewed).

Kotlin, conversely, provides coroutines: a built-in mechanism, at the language level, for executing asynchronous tasks.
Twenty-one developers highlighted coroutines as one of the best features that Kotlin provides.
As one developer told us:
\quotes{In my opinion, coroutines allow developers to write easy to read and concise code, that can be read top-down, like a book. I believe that's a big advantage}.
Another developer remarked: \quotes{coroutines were the cleanest way to implement my complex workflows}.

Eight developers mentioned that they had replaced RxJava with coroutines.
One told us that: 
\quotes{the code with RxJava looks `hacky' and it quickly becomes a mess}, 
and another developer:
\quotes{Coroutines are also absolutely great. They have allowed me to drop RxJava, which although very useful, was annoying to work with in complicated situations. Being able to write asynchronous code in direct style is wonderful}.

As well as all the positive perceptions collected, 
just one developer told us that, although he uses coroutines, he finds they entail a \quotes{lot of complexity}.

{\bf Implication:} Android developers can simplify the code for their asynchronous tasks by using the language-level supported feature called `coroutines'.
Beyond that, further research may need to compare coroutines with other concurrency mechanisms (e.g., RxJava) in terms of other dimensions such as performance.

\section{Threats to Validity}
\label{sec:tov}

$\; \; $ 
\point{The applications studied might not be representative of open-source applications}
we studied 374 open-source Kotlin Android applications coming from two datasets.
There is a risk that these apps might not be representative of open-source Android applications written in Kotlin.
However, we apply a methodology to capture and analyze the largest number of open-source Android applications published on apps markets.

\point{Open-source apps might not be representative of all Android apps}
our study focuses on publicly available open-source apps because we need a repository with visible source code to detect migrations.
Thus, we cannot analyze non-open-source apps published on app stores. 
For further analysis, the list of open-source apps analyzed is available in our appendix~\cite{appendix}.

\point{Heuristic for detecting migrations}
we identify developers that migrate code by using one particular heuristic (See \ref{sec:heuristics_detection_commits}).
There could be other manners of migrating code that our heuristic cannot detect.
Therefore, more research is required on code migrations, e.g., to define a taxonomy of code migrations.

\point{Accuracy of \miga{}}
there might be a risk that a bug in our tool affects the results we present.
We manually inspected a sample of migration commits detected by the tool and did not find any anomalies. 
Our tool is publicly available~\cite{migatool}. Thus researchers can inspect its code, use and extend it.

\point{Sample}
qualitative studies such as ours
are performed by using a representative sample of the population under study~\cite{experimentationSE2012}.
The \emph{target} population corresponds to developers that have migrated code from Java to Kotlin in Android apps.
We create a \emph{sampling frame} by retrieving developers (detected in open-source code repositories) that have migrated Java code to Kotlin.
Then, we send emails to \emph{all} of them.
It might be the case that the developers that we contact and that reply to us represent neither the Android community nor the Kotlin community.
To minimize this risk, we analyze \emph{all} applications from two of the largest Android open-source app repositories: F-Droid and AndroidTimeMachine~\cite{Geiger2018:data}.

\point{Bias in responses}
there is a risk that the responses given by the developers are biased:
our study captures the experience of developers that successfully migrated and integrated Kotlin code into their codebases.
Our \new{methodological approach} for finding developers is based on commit analysis. 
Therefore, we cannot identify developers who tried Kotlin, but later decided not to adopt it (and thus never committed any Kotlin code into their repositories).
This would require contacting all developers that had never committed Kotlin code, something which is unfeasible.

\point{
{Research design}}
We based our {research} design on an initial question to reduce the costs to participants. 
There is a risk that the responses we received did not completely capture participants' experiences with Kotlin (e.g., a developer forgot to mention one of her/his favorites Kotlin features).
To mitigate a developer omitting information, we tried to contact as large a number as possible of developers that migrated code on open-source Android apps.

\point{Reliability of answers}
some answers might not reliably reflect the experiences of the developers that wrote them. 
As we have a considerable number of answers from developers, having a few imprecise or incorrect answers does not invalidate the overall results.

\point{Verification of qualitative research}
our study follows a qualitative research method that attempts to interpret a phenomenon (i.e., migrations) based on explanations that people (i.e., developers) bring forward~\cite{experimentationSE2012}. 
We follow a verification mechanism used during qualitative research to ensure the reliability and validity of our study.
For instance, the open and axial coding (see Section \ref{sec:methodology:recolections}) applied to mine codes from the qualitative data (the emails) was performed by one of the authors of this paper and later verified and validated by the other author.

\point{Mining the grey-literature}
in this paper, we analyze some articles from the grey literature on Kotlin adoption and migration.
Our study neither aims to nor claim to be a systematic literature review on those topics.
Consequently, there may be, for instance, migration problems that this research does not report, but that have been reported by articles from the grey literature (or other artifacts such as Audio-Video media \cite{GAROUSI2019101})  that we did not inspect.
Nevertheless, in order to be transparent about the process, in our appendix~\cite{appendix} we report the query we use and the links to analyzed sites.

\section{Discussion and Future work}
\label{sec:discussion}

\subsection{
Additional codes from the grey literature}
\label{sec:differencesgreyliterature}

We list some codes we mine from the grey literature that are not discussed by developers in our primary study.
These points could represent future research opportunities, in addition to those detailed in Section \ref{sec:futurework} that arose from the interviews.

\point{Kotlin is slower to compile}
Several posts have tested the compilation time of Kotlin code (\cite{Heath20LinkSwitch, daga2018LinkJavaVsKotlin, Kust2017LinkKotlinFeatures, Rattra2020, Hiral2018}).
For example, Daga~\cite{daga2018LinkJavaVsKotlin} reports a difference of 13\% on the compilation time w.r.t. Java. 
Wilson~\cite{Wilson2019Metrics} reports that, after migrating a library named OkHttp's from Java to Kotlin, the compilation time went from 2.4 to 10.2 seconds.
Engineers from Uber~\cite{Fernandes2019UberPerformance} found that the Kotlin's new type inference system adds an overhead of 8\% in total compilation time.

\point{Kotlin increases the size of the APK}
Some of the grey literature posts mention the increase in the size of the APKs (the package file format used by Android, which contains compiled code) when Kotlin is used.
For example, 
Wilson~\cite{Wilson2019Metrics} reports that, after migrating its library named OkHttp’s from Java to Kotlin, the binary (APK) size increased by 60\%.
Daga~\cite{daga2018LinkJavaVsKotlin} reports an average increases of 800 KB and 1 MB, and 
Kust~\cite{Kust2017LinkKotlinFeatures} reports an approximate 300 KB increase.

\point{Absence of checked exceptions in Kotlin}
unlike Java, Kotlin does not have \emph{checked} exceptions (i.e., types of exceptions that must either be caught or declared in the method in which it is thrown).
Articles from the grey literature highlight some benefits of getting by without them: 
Daga~\cite{daga2018LinkJavaVsKotlin} states \quotes{Kotlin does away with them to aid code conciseness}, and 
Gour~\cite{Gour2020} states \quotes{it can minimize the verbosity and improve type safety}.

\point{Problems with Wildcard-types}
Kotlin has neither primitive (raw) types nor Wildcard-types (i.e., the symbol `?' in Java which represents an unknown type in generic programming)~\cite{Hiral2018}.
Kotlin code that uses wildcards and that was generated by the auto-converter tool can produce runtime failures according to \cite{AbdelAziz2020MigEnterprise}. 
This means developers must revise and modify, as we mention in Section \ref{sec:autoconverter}, the output from the auto-converter tool.

\point{Problems with annotations} 
A post by Thornsby \cite{Thornsby2019LinkJavaVsKotlinKeyDiff} mentions that one of the differences between developing in Java and Kotlin is the use of KATP, the official tool for  processing annotations.
A post by Hofmann~\cite{Hofmann2017Migration} states that a problem of adopting Kotlin was related to KATP, which produced build errors when it was integrated into other libraries such as Dagger\footnote{Dagger: \url{https://dagger.dev/}}, a dependency injection library. 
That integration was also reported as problematic by~\cite{Papadopoulos2018Migration}.
Related to this, engineers from Uber found that using KATP adds a $\sim$95\% overhead ~\cite{Fernandes2019UberPerformance}.

\point{Performance issues}
Athaydes~\cite{Athaydes2017Benchmark} and BeyksZ~\cite{Beyls2017HidenCost} discuss and measure the runtime overhead introduced by Kotlin features. 
The former shows that 8 out of 11 evaluated features have an overhead lower than 5\%,  withthe remaining 3 (\begin{inparaenum}[]
\item Varargs + Spread Operator, 
\item Delegate Properties and 
\item Ranges)
\end{inparaenum} having an overhead that is higher than 10\%.
We consider that more research on the performance of Kotlin is needed.

\subsection{Research Challenges}
\label{sec:directions}
\label{sec:futurework}

Wohlin et al. \cite{experimentationSE2012} indicate that an exploratory study, like the one conducted in this paper, can be used as pre-study for a more thorough investigation, and may provide new possibilities that could be analyzed and should therefore be followed up in a more focused or thorough survey. 
In this section, we present some new possibilities that could be exploited by the research community.

\point{Auto-converter tools}
The study we conduct shows that developers used the auto-converter code and modified the output code to make it more idiomatic.
The automated-generated code is like \quotes{code written with Kotlin syntax but still keeping Java style}.
Future research could define new automated code conversion approaches and tools (like the initial work from \cite{CATA2019:Vetting}) for generating idiomatic code, and consequently, attempting to reduce the modifications that developers have to make.

\point{Refactoring Kotlin code after migration}
A challenge for researchers is to define approaches and tools that propose \emph{refactorings} on Kotlin applications after the migration has been (partially or fully) carried out.
We envision two directions.
First, for identifying and potentially automatically removing {code} with technical-debt items, such as code smells and code patterns~\cite{avgeriou_et_al:DR:2016:6693}, related to Kotlin code.
Existing tools target Java code (e.g., JDeodorant~\cite{Tsantalis2008JDeodorant}) but, to our knowledge, none targets Kotlin code.
Secondly, we envision refactorings that make Kotlin code (both that  written by humans or generated by a tool) more idiomatic. 
For example, one could propose using new Kotlin features (e.g., \emph{smart casts}) that had not previously been used by the developer.

\point{Supporting migration activity}
A potential research direction is to define guidelines, approaches, and tools that help developers migrate their apps.
Gradual migrations are convenient for migrating large applications (e.g., Duolingo~\cite{duolingo_migration2020}) because they allow developers and companies to migrate some parts of their apps while simultaneously being able to evolve other parts, add new features and release new versions.

Gradual (or iterative) migration has already been studied on legacy applications~(e.g., \cite{DeLucia2008DevelopingLegacy}). It consists of first decomposing the legacy apps~\cite{Brodie1995Legacy} and then integrating the decomposed parts by using \emph{wrappers} (e.g.,~\cite{CANFORA200099}). 
Migration from Java to Kotlin does not need such wrappers, as both languages are interoperable. 
In this scenario, a developer needs to define a \emph{migration strategy} that defines consecutive migration steps, and for each of them, a subset of components or code files to be migrated.
This could be challenging, especially on large applications: the wrong selection of files to be migrated could increase the migration effort due to emerging errors~\cite{AbdelAziz2020MigEnterprise}. 
We envision the definition of guidelines, approaches, and tools that help developers define a migration strategy and select the correct files, packages, or components to be migrated in one particular step to minimize the migration effort. An initial experiment on file recommendation for migration was performed by~\cite{mateus2021experiencebased}.

\point{Estimating the cost of migration}
One of the main problems about migrations reported by our study and literature on migration (e.g., \cite{Khadka2014LegacyModernization}) is unfinished migrations due to the lack of time or resources.
Models for estimating the cost and effort of migration could help developers and companies forecast these issues.
A challenge for researchers that focus on migration is to define models that estimate the cost of {migrating} Android apps to Kotlin based on existing estimation models~\cite{Jorgensen2007SurveyCost}.

\point{Kotlin code comprehensibility} 
in our study, some developers mentioned that the compact syntax of Kotlin affected the readability of the code.
A research direction could focus on empirically studying the comprehensibility of Kotlin code.
To our knowledge, there is no previous work on this.
Future work could focus on defining tools that would attempt to automatically measure code comprehensibility, and to point out sections of code that are difficult to comprehend and should therefore be refactored~\cite{Wyrich2020Comprehensibility}.

\subsection{\new{Research Ethics}}
\label{sec:ethics}

Since our \new{study} involved human subjects, as recommended by \cite{experimentationSE2012}, we considered ethical aspects when we designed the study:
\begin{inparaenum}[1)]
\item we informed subjects of the research project's title and goals,
\item we mentioned our affiliations, and provided links to our professional websites,
\item we explained that participation would help us write a research paper,
\item we guaranteed the confidentiality of the information obtained, i.e., we did not share the answers,
\item every quote included in this research is anonymized, and 
\item \new{names of developers have been codified in order to protect their privacy}.

\end{inparaenum}
Nevertheless, we consider that developers that responded to this study (as well other developers we did not contact) will benefit from this research in the long run because, for instance, we enumerate and detail the various different problems that they encountered during the migration process, allowing the research community to focus on providing solutions to them.

\section{Related Work}
\label{sec:rw}

\subsection{Adoption of Kotlin}
Oliveira et al. \cite{OliveiraSANER20p206} conducted a study to understand how developers deal with the adoption of Kotlin as an official language for Android development, their perception of the advantages and disadvantages of it, and the most common problems they face.
Our work has a narrower goal than their study: to study the migration of Android applications from Java to Kotlin.
Moreover, the surveys conducted have two main differences:
\begin{inparaenum}[1)]
\item the goal: to obtain the experience of Kotlin adoption \cite{OliveiraSANER20p206} vs experience of migration (us),
\item the size: 7 interviewees \cite{OliveiraSANER20p206} vs \nremailreplies{} developers that migrated code (us).
\end{inparaenum}

\subsection{Surveys on legacy system migration and software modernization}
Torchiano et al. \cite{Torchiano2008MigrationIndustry} conducted a survey that identified the state-of-the-practice of 59 software migration projects in Italian industries.
There are three main differences to our work.
The study differs from our work in three main ways.
First, the scope: \cite{Torchiano2008MigrationIndustry} focuses on industrial applications, and we focus on open-source apps.
Second, \cite{Torchiano2008MigrationIndustry} does not focus on a particular technology, and we focus on Kotlin and Android.
Third, the contacted subjects: \cite{Torchiano2008MigrationIndustry} targets different profiles (39.7\% were software developers); we focus exclusively  on developers.

Khadka et al. \cite{Khadka2014LegacyModernization} conducted an exploratory study to discover new perspectives, insights, drivers of modernization and challenges with respect to legacy systems in industry.
It consisted of semi-structured interviews with 26 industrial practitioners.
The findings were then validated through a separate structured survey that involved 198 professionals.
The differences: our study focuses on open-source apps (\cite{Khadka2014LegacyModernization} focuses on industrial projects) migrated from Java to Kotlin on Android (\cite{Khadka2014LegacyModernization} does not focus on particular technologies).

Other works have focused on the migration of specific technologies by conducting surveys.
For example, Razavian and Lago~\cite{Razavian2021SurveySOA} conducted a survey in seven companies to understand the migration approaches from legacy systems to service-oriented applications (SOA).

\subsection{Empirical studies on Kotlin code}
Gois Mateus and Martinez \cite{GoisMateus2019} found that 11\% of the studied open-source applications from Android have Kotlin code.
They also found that the adoption of Kotlin increases the quality of applications, expressed in terms of the presence of code smells.
Flauzino et al. \cite{Flauzino:2018:YSS} studied 100 repositories of programs containing Java or Kotlin code. They found that, on average, Kotlin programs have fewer code smells than Java programs.
Gois and Martinez \cite{Gois2020Features} have studied the adoption of features introduced by Kotlin.
Ardito et al. \cite{Ardito2020KotlinJava} conducted a study with undergraduate students to assess the assumed advantages of Kotlin over Java in the context of Android development and maintenance.
They found evidence that the adoption of Kotlin led to more compact code. 
These findings from students are aligned to our findings derived from developers.

\subsection{Studies on library migration}
Other works focus on empirical studies of migrating of libraries (e.g., \cite{Teyton:2014:SLM, Kula:2018:DUL, hora2018sqj}) and recommendations on library migrations (e.g., \cite{Teyton2013Discovering, Alrubaye:2019:UIR}).
For instance, Salza et al. \cite{Salza:2018:DUT} focus on studying the migration of third-party libraries in mobile apps. 
They find that mobile developers rarely update their apps.
In this paper, we exclusively focus on the migration of source code. 

\section{Conclusion}
\label{sec:conclusion}
In this work, we conducted a qualitative study that gathered data from \nremailreplies{} Android developers who migrated code from Java to Kotlin to ask them {Why} they had migrated that code.
Developers found using Kotlin a way to
\begin{inparaenum}[a)]
\item use modern features not available in  Java in their Android applications (e.g., extending function) and to
\item have safer, shorter and less verbose code. 
\end{inparaenum}

In addition, we defined different research directions extracted from our study based on the difficulties  developers had encountered when they migrated code.
In future work, to help developers to overcome the difficulties they currently have when they adopt Kotlin and migrate code from Java to Kotlin,
we plan to work on:
\begin{inparaenum}[1)]
\item defining an approach that translates code from Java to Kotlin and produces more idiomatic code;
\item defining an approach that refactors Kotlin code to remove code smells, to improve code comprehensibility, and to obtain more idiomatic code; 
\item helping developers to migrate large applications by presenting migration strategies (for example, suggesting files and components to migrate first);
and 
\item improving the testing of the migrated parts of systems under migration, as well as the unmigrated parts that interact with the migrated parts.
\end{inparaenum}

\vspace{-0.4cm}

\section*{Acknowledgements}

We would like to thank the \nremailreplies{} developers that took the time to answer our questions, and Prof. Lara Maestripieri for advising us about how to conduct qualitative studies, verifying the methods we applied and for revising our manuscript. 

\bibliography{references}

\begin{thebibliography}{10}
\providecommand{\url}[1]{#1}
\csname url@samestyle\endcsname
\providecommand{\newblock}{\relax}
\providecommand{\bibinfo}[2]{#2}
\providecommand{\BIBentrySTDinterwordspacing}{\spaceskip=0pt\relax}
\providecommand{\BIBentryALTinterwordstretchfactor}{4}
\providecommand{\BIBentryALTinterwordspacing}{\spaceskip=\fontdimen2\font plus
\BIBentryALTinterwordstretchfactor\fontdimen3\font minus
  \fontdimen4\font\relax}
\providecommand{\BIBforeignlanguage}[2]{{%
\expandafter\ifx\csname l@#1\endcsname\relax
\typeout{** WARNING: IEEEtran.bst: No hyphenation pattern has been}%
\typeout{** loaded for the language `#1'. Using the pattern for}%
\typeout{** the default language instead.}%
\else
\language=\csname l@#1\endcsname
\fi
#2}}
\providecommand{\BIBdecl}{\relax}
\BIBdecl

\bibitem{mobileshare}
\BIBentryALTinterwordspacing
IDC. (2017) Smartphone os market share, 2017 q1. [Online]. Available:
  \url{https://www.idc.com/promo/smartphone-market-share/}
\BIBentrySTDinterwordspacing

\bibitem{Shafirov2017LinkKotlinOfficialInAndroidJetBrains}
\BIBentryALTinterwordspacing
M.~Shafirov. (2017, May) Kotlin on android. now official. [Online]. Available:
  \url{https://blog.jetbrains.com/kotlin/2017/05/kotlin-on-android-now-official/}
\BIBentrySTDinterwordspacing

\bibitem{Coppola:2019:CTK}
\BIBentryALTinterwordspacing
R.~Coppola, L.~Ardito, and M.~Torchiano, ``Characterizing the transition to
  kotlin of android apps: A study on f-droid, play store, and github,'' in
  \emph{Proceedings of the 3rd ACM SIGSOFT International Workshop on App Market
  Analytics}, ser. WAMA 2019.\hskip 1em plus 0.5em minus 0.4em\relax New York,
  NY, USA: ACM, 2019, pp. 8--14. [Online]. Available:
  \url{http://doi.acm.org/10.1145/3340496.3342759}
\BIBentrySTDinterwordspacing

\bibitem{OliveiraSANER20p206}
V.~Oliveira, L.~Teixeira, and F.~Ebert, ``On the adoption of kotlin on android
  development: A triangulation study,'' in \emph{Proc.\ SANER}.\hskip 1em plus
  0.5em minus 0.4em\relax IEEE, 2020, pp. 206--216.

\bibitem{Brodie1995Legacy}
M.~L. Brodie and M.~Stonebraker, \emph{Legacy Information Systems Migration:
  Gateways, Interfaces, and the Incremental Approach}.\hskip 1em plus 0.5em
  minus 0.4em\relax San Francisco, CA, USA: Morgan Kaufmann Publishers Inc.,
  1995.

\bibitem{Bisbal1997Migration}
J.~{Bisbal}, D.~{Lawless}, {Bing Wu}, and J.~{Grimson}, ``Legacy information
  systems: issues and directions,'' \emph{IEEE Software}, vol.~16, no.~5, pp.
  103--111, Sep. 1999.

\bibitem{Colosimo2009Evaluatinglegacy}
\BIBentryALTinterwordspacing
M.~Colosimo, A.~D. Lucia, G.~Scanniello, and G.~Tortora, ``Evaluating legacy
  system migration technologies through empirical studies,'' \emph{Information
  and Software Technology}, vol.~51, no.~2, pp. 433 -- 447, 2009. [Online].
  Available:
  \url{http://www.sciencedirect.com/science/article/pii/S0950584908000694}
\BIBentrySTDinterwordspacing

\bibitem{GoisMateus2019}
\BIBentryALTinterwordspacing
B.~G{\'o}is~Mateus and M.~Martinez, ``An empirical study on quality of android
  applications written in kotlin language,'' \emph{Empirical Software
  Engineering}, Jun 2019. [Online]. Available:
  \url{https://doi.org/10.1007/s10664-019-09727-4}
\BIBentrySTDinterwordspacing

\bibitem{saunders2009research}
M.~Saunders, P.~Lewis, and A.~Thornhill, \emph{Research methods for business
  students}.\hskip 1em plus 0.5em minus 0.4em\relax Pearson education, 2009.

\bibitem{migatool}
\BIBentryALTinterwordspacing
Miga tool. [Online]. Available: \url{https://github.com/UPHF/migA}
\BIBentrySTDinterwordspacing

\bibitem{coming2019}
M.~{Martinez} and M.~{Monperrus}, ``Coming: A tool for mining change pattern
  instances from git commits,'' in \emph{2019 IEEE/ACM 41st International
  Conference on Software Engineering: Companion Proceedings (ICSE-Companion)},
  May 2019, pp. 79--82.

\bibitem{silverman2013doing}
D.~Silverman, \emph{Doing qualitative research: A practical handbook}.\hskip
  1em plus 0.5em minus 0.4em\relax Sage, 2013.

\bibitem{experimentationSE2012}
C.~Wohlin, P.~Runeson, M.~Hst, M.~C. Ohlsson, B.~Regnell, and A.~Wessln,
  \emph{Experimentation in Software Engineering}.\hskip 1em plus 0.5em minus
  0.4em\relax Springer Publishing Company, Incorporated, 2012.

\bibitem{patton1990qualitative}
M.~Q. Patton, \emph{Qualitative evaluation and research methods}.\hskip 1em
  plus 0.5em minus 0.4em\relax SAGE Publications, inc, 1990.

\bibitem{appendix}
\BIBentryALTinterwordspacing
Appendix. [Online]. Available:
  \url{https://github.com/UPHF/kotlin_migration_experiment}
\BIBentrySTDinterwordspacing

\bibitem{glaser2017discovery}
B.~G. Glaser and A.~L. Strauss, \emph{Discovery of grounded theory: Strategies
  for qualitative research}.\hskip 1em plus 0.5em minus 0.4em\relax Routledge,
  2017.

\bibitem{cohen2002research}
L.~Cohen, L.~Manion, and K.~Morrison, \emph{Research methods in
  education}.\hskip 1em plus 0.5em minus 0.4em\relax routledge, 2002.

\bibitem{Geiger2018:data}
\BIBentryALTinterwordspacing
F.-X. Geiger, I.~Malavolta, L.~Pascarella, F.~Palomba, D.~Di~Nucci, and
  A.~Bacchelli, ``A graph-based dataset of commit history of real-world android
  apps,'' in \emph{Proceedings of the 15th International Conference on Mining
  Software Repositories}, ser. MSR '18.\hskip 1em plus 0.5em minus 0.4em\relax
  New York, NY, USA: ACM, 2018, pp. 30--33. [Online]. Available:
  \url{http://doi.acm.org/10.1145/3196398.3196460}
\BIBentrySTDinterwordspacing

\bibitem{Kovalenko2018Branches}
\BIBentryALTinterwordspacing
V.~Kovalenko, F.~Palomba, and A.~Bacchelli, ``Mining file histories: Should we
  consider branches?'' in \emph{Proceedings of the 33rd ACM/IEEE International
  Conference on Automated Software Engineering}, ser. ASE 2018.\hskip 1em plus
  0.5em minus 0.4em\relax New York, NY, USA: Association for Computing
  Machinery, 2018, p. 202–213. [Online]. Available:
  \url{https://doi.org/10.1145/3238147.3238169}
\BIBentrySTDinterwordspacing

\bibitem{Coelho2015ExceptionsAndroid}
R.~Coelho, L.~Almeida, G.~Gousios, and A.~van Deursen, ``Unveiling exception
  handling bug hazards in android based on github and google code issues,'' in
  \emph{Proceedings of the 12th Working Conference on Mining Software
  Repositories}, ser. MSR ’15.\hskip 1em plus 0.5em minus 0.4em\relax IEEE
  Press, 2015, p. 134–145.

\bibitem{Butterworth2020LinkMigration}
\BIBentryALTinterwordspacing
J.~Butterworth. (2020, Feb) Migrating to kotlin—what to look out for.
  [Online]. Available:
  \url{https://engineering.autotrader.co.uk/2020/02/21/migrating-to-kotlin-what-to-look-out-for.html}
\BIBentrySTDinterwordspacing

\bibitem{Caneco2017LinkMigrating}
\BIBentryALTinterwordspacing
N.~Caneco. (2017, May) Migrating from java to kotlin: the easy way. [Online].
  Available:
  \url{https://engineering.talkdesk.com/migrating-from-java-to-kotlin-the-easy-way-37b25a379d72}
\BIBentrySTDinterwordspacing

\bibitem{daga2018LinkJavaVsKotlin}
\BIBentryALTinterwordspacing
M.~Daga. (2018, May) Java vs kotlin: Which programming language is better for
  android developers? [Online]. Available:
  \url{https://dzone.com/articles/java-vs-kotlin-which-programming-language-is-bette}
\BIBentrySTDinterwordspacing

\bibitem{Sommerhoff2015LinkKotlin}
\BIBentryALTinterwordspacing
P.~Sommerhoff. (2015, Dec) Kotlin for java developers: 10 features you will
  love about kotlin. [Online]. Available:
  \url{http://petersommerhoff.com/dev/kotlin/kotlin-for-java-devs/}
\BIBentrySTDinterwordspacing

\bibitem{Heath20LinkSwitch}
\BIBentryALTinterwordspacing
N.~Heath. (2019, May) Should android devs switch from java to kotlin? here's
  google's advice on swapping programming languages. [Online]. Available:
  \url{https://www.techrepublic.com/article/should-android-devs-switch-from-java-to-kotlin-heres-googles-advice-on-swapping-programming/}
\BIBentrySTDinterwordspacing

\bibitem{Sommerhoff2020LinkJavaVsKotlin}
\BIBentryALTinterwordspacing
P.~Sommerhoff. (2020, Feb) Kotlin vs. java: 9 benefits of kotlin for your
  business. [Online]. Available:
  \url{https://blog.udemy.com/kotlin-vs-java-9-benefits-of-kotlin-for-your-business/}
\BIBentrySTDinterwordspacing

\bibitem{Rishabh2017LinkKotlinAndroid}
\BIBentryALTinterwordspacing
R.~Software. (2017, Sept) Android development using kotlin: Streamline the
  development workflow. [Online]. Available:
  \url{https://www.rishabhsoft.com/blog/kotlin-for-android-development}
\BIBentrySTDinterwordspacing

\bibitem{Thornsby2019LinkJavaVsKotlinKeyDiff}
\BIBentryALTinterwordspacing
J.~Thornsby. (2019, Oct) Kotlin vs java for android: key differences. [Online].
  Available: \url{https://www.androidauthority.com/kotlin-vs-java-783187/}
\BIBentrySTDinterwordspacing

\bibitem{Hiral2018}
\BIBentryALTinterwordspacing
H.~Atha. (2018, Aug) Java vs kotlin – which should you choose for android
  development. [Online]. Available:
  \url{https://www.moveoapps.com/blog/java-vs-kotlin/}
\BIBentrySTDinterwordspacing

\bibitem{duolingo_website}
\BIBentryALTinterwordspacing
(2020, April) Press duolingo. [Online]. Available:
  \url{https://www.duolingo.com/press}
\BIBentrySTDinterwordspacing

\bibitem{Gois2020Features}
\BIBentryALTinterwordspacing
B.~G. Mateus and M.~Martinez, ``On the adoption, usage and evolution of kotlin
  features in android development,'' in \emph{Proceedings of the 14th ACM /
  IEEE International Symposium on Empirical Software Engineering and
  Measurement (ESEM)}, ser. ESEM '20.\hskip 1em plus 0.5em minus 0.4em\relax
  New York, NY, USA: Association for Computing Machinery, 2020. [Online].
  Available: \url{https://doi.org/10.1145/3382494.3410676}
\BIBentrySTDinterwordspacing

\bibitem{Kust2017LinkKotlinFeatures}
\BIBentryALTinterwordspacing
I.~Kušt. (2017) Ten kotlin features to boost android development. [Online].
  Available:
  \url{https://www.toptal.com/android/kotlin-boost-android-development}
\BIBentrySTDinterwordspacing

\bibitem{Gour2020}
\BIBentryALTinterwordspacing
R.~Gour. (2020, March) Why you must switch from java to kotlin for android
  development? [Online]. Available:
  \url{https://codeburst.io/why-you-must-switch-from-java-to-kotlin-for-android-development-6cf179d16dc7}
\BIBentrySTDinterwordspacing

\bibitem{AbdelAziz2020MigEnterprise}
\BIBentryALTinterwordspacing
M.~Abdelaziz. (2020) Migrating java enterprise apps to kotlin. [Online].
  Available: \url{https://vaadin.com/learn/tutorials/migrate-to-kotlin}
\BIBentrySTDinterwordspacing

\bibitem{strumenta2020}
\BIBentryALTinterwordspacing
Strumenta. (2020) Migration of java applications to kotlin. [Online].
  Available: \url{https://superkotlin.com/java-to-kotlin-migrations/}
\BIBentrySTDinterwordspacing

\bibitem{Khadka2014LegacyModernization}
\BIBentryALTinterwordspacing
R.~Khadka, B.~V. Batlajery, A.~M. Saeidi, S.~Jansen, and J.~Hage, ``How do
  professionals perceive legacy systems and software modernization?'' in
  \emph{Proceedings of the 36th International Conference on Software
  Engineering}, ser. ICSE 2014.\hskip 1em plus 0.5em minus 0.4em\relax New
  York, NY, USA: Association for Computing Machinery, 2014, p. 36–47.
  [Online]. Available: \url{https://doi.org/10.1145/2568225.2568318}
\BIBentrySTDinterwordspacing

\bibitem{Trehan2019Migration}
\BIBentryALTinterwordspacing
K.~Trehan. (2019, Jun) Kotlin migration @pepperfry. [Online]. Available:
  \url{https://medium.com/pepperfry-tech/kotlin-migration-pepperfry-part-one-motivation-b05102bf0e7a}
\BIBentrySTDinterwordspacing

\bibitem{Wilson2019Metrics}
\BIBentryALTinterwordspacing
J.~Wilson. (2019, May) Metrics for okhttp’s kotlin upgrade. [Online].
  Available:
  \url{https://publicobject.com/2019/05/13/metrics-for-okhttps-kotlin-upgrade/}
\BIBentrySTDinterwordspacing

\bibitem{Fernandes2019UberPerformance}
\BIBentryALTinterwordspacing
E.~Fernandes, T.~Machado, T.~Nguyen, and Z.~Sweers. (2019, Apr) Measuring
  kotlin build performance at uber. [Online]. Available:
  \url{https://eng.uber.com/measuring-kotlin-build-performance/}
\BIBentrySTDinterwordspacing

\bibitem{Vavra2017}
\BIBentryALTinterwordspacing
D.~Vávra. (2017, Jun) How to remove all !! from your kotlin code. [Online].
  Available:
  \url{https://android.jlelse.eu/how-to-remove-all-from-your-kotlin-code-87dc2c9767fb}
\BIBentrySTDinterwordspacing

\bibitem{Baxter2017}
\BIBentryALTinterwordspacing
B.~Baxter. (2017, Jun) Lessons learned while converting to kotlin with android
  studio. [Online]. Available:
  \url{https://medium.com/androiddevelopers/lessons-learned-while-converting-to-kotlin-with-android-studio-f0a3cb41669}
\BIBentrySTDinterwordspacing

\bibitem{Verdecchia2019Guidelines}
R.~{Verdecchia}, I.~{Malavolta}, and P.~{Lago}, ``Guidelines for architecting
  android apps: A mixed-method empirical study,'' in \emph{2019 IEEE
  International Conference on Software Architecture (ICSA)}, 2019, pp.
  141--150.

\bibitem{GAROUSI2019101}
\BIBentryALTinterwordspacing
V.~Garousi, M.~Felderer, and M.~V. Mäntylä, ``Guidelines for including grey
  literature and conducting multivocal literature reviews in software
  engineering,'' \emph{Information and Software Technology}, vol. 106, pp. 101
  -- 121, 2019. [Online]. Available:
  \url{http://www.sciencedirect.com/science/article/pii/S0950584918301939}
\BIBentrySTDinterwordspacing

\bibitem{Rattra2020}
\BIBentryALTinterwordspacing
S.~Rattra. (2020, May) Kotlin vs java - android development. [Online].
  Available:
  \url{https://hackernoon.com/kotlin-vs-java-android-development-qh6z329j}
\BIBentrySTDinterwordspacing

\bibitem{Hofmann2017Migration}
\BIBentryALTinterwordspacing
P.~Hofmann. (2017, Nov) Migrating an android app from java to kotlin. [Online].
  Available:
  \url{https://medium.com/monsterculture/moving-from-java-to-kotlin-on-android-f60c593c39f8}
\BIBentrySTDinterwordspacing

\bibitem{Papadopoulos2018Migration}
\BIBentryALTinterwordspacing
A.~Papadopoulos. (2018, Jun) Effective migration to kotlin on android.
  [Online]. Available:
  \url{https://android.jlelse.eu/effective-migration-to-kotlin-on-android-cfb92bfaa49b}
\BIBentrySTDinterwordspacing

\bibitem{Athaydes2017Benchmark}
\BIBentryALTinterwordspacing
R.~Athaydes. (2017, Oct) Kotlin's hidden costs - benchmarks. [Online].
  Available:
  \url{https://sites.google.com/a/athaydes.com/renato-athaydes/posts/kotlinshiddencosts-benchmarks}
\BIBentrySTDinterwordspacing

\bibitem{Beyls2017HidenCost}
\BIBentryALTinterwordspacing
C.~Beyls. (2017, Jul) Exploring kotlin’s hidden costs. [Online]. Available:
  \url{https://bladecoder.medium.com/exploring-kotlins-hidden-costs-part-1-fbb9935d9b62}
\BIBentrySTDinterwordspacing

\bibitem{CATA2019:Vetting}
\BIBentryALTinterwordspacing
C.~Courtney and M.~Neilsen, ``Vetting anti-patterns in java to kotlin
  translation,'' in \emph{Proceedings of 34th International Conference on
  Computers and Their Applications}, ser. EPiC Series in Computing, G.~Lee and
  Y.~Jin, Eds., vol.~58.\hskip 1em plus 0.5em minus 0.4em\relax EasyChair,
  2019, pp. 191--202. [Online]. Available:
  \url{https://easychair.org/publications/paper/Jdw6}
\BIBentrySTDinterwordspacing

\bibitem{avgeriou_et_al:DR:2016:6693}
\BIBentryALTinterwordspacing
P.~Avgeriou, P.~Kruchten, I.~Ozkaya, and C.~Seaman, ``{Managing Technical Debt
  in Software Engineering (Dagstuhl Seminar 16162)},'' \emph{Dagstuhl Reports},
  vol.~6, no.~4, pp. 110--138, 2016. [Online]. Available:
  \url{http://drops.dagstuhl.de/opus/volltexte/2016/6693}
\BIBentrySTDinterwordspacing

\bibitem{Tsantalis2008JDeodorant}
N.~{Tsantalis}, T.~{Chaikalis}, and A.~{Chatzigeorgiou}, ``Jdeodorant:
  Identification and removal of type-checking bad smells,'' in \emph{2008 12th
  European Conference on Software Maintenance and Reengineering}, 2008, pp.
  329--331.

\bibitem{duolingo_migration2020}
\BIBentryALTinterwordspacing
A.~Chaidarun. (2020, April) Migrating duolingo’s android app to 100\% kotlin.
  [Online]. Available:
  \url{https://blog.duolingo.com/migrating-duolingos-android-app-to-100-kotlin/}
\BIBentrySTDinterwordspacing

\bibitem{DeLucia2008DevelopingLegacy}
A.~De~Lucia, R.~Francese, G.~Scanniello, and G.~Tortora, ``Developing legacy
  system migration methods and tools for technology transfer,'' \emph{Softw.
  Pract. Exper.}, vol.~38, no.~13, p. 1333–1364, Nov. 2008.

\bibitem{CANFORA200099}
\BIBentryALTinterwordspacing
G.~Canfora, A.~Cimitile, A.~{De Lucia}, and G.~A. {Di Lucca}, ``Decomposing
  legacy programs: a first step towards migrating to client–server
  platforms,'' \emph{Journal of Systems and Software}, vol.~54, no.~2, pp. 99
  -- 110, 2000, special Issue on Software Maintenance. [Online]. Available:
  \url{http://www.sciencedirect.com/science/article/pii/S0164121200000303}
\BIBentrySTDinterwordspacing

\bibitem{mateus2021experiencebased}
\BIBentryALTinterwordspacing
B.~G. Mateus, C.~Kolski, and M.~Martinez, ``An experience-based recommendation
  system to support migrations of android applications from java to kotlin,''
  ArXiV, Tech. Rep. 2103.09728, March 2021. [Online]. Available:
  \url{http://arxiv.org/abs/2103.09728}
\BIBentrySTDinterwordspacing

\bibitem{Jorgensen2007SurveyCost}
M.~{Jorgensen} and M.~{Shepperd}, ``A systematic review of software development
  cost estimation studies,'' \emph{IEEE Transactions on Software Engineering},
  vol.~33, no.~1, pp. 33--53, 2007.

\bibitem{Wyrich2020Comprehensibility}
M.~Wyrich, A.~Preikschat, D.~Graziotin, and S.~Wagner, ``The mind is a powerful
  place: How showing code comprehensibility metrics influences code
  understanding,'' in \emph{Proceedings of the 43rd International Conference on
  Software Engineering (ICSE '21)}, 2021.

\bibitem{Torchiano2008MigrationIndustry}
M.~{Torchiano}, M.~{Di Penta}, F.~{Ricca}, A.~{De Lucia}, and F.~{Lanubile},
  ``Software migration projects in italian industry: Preliminary results from a
  state of the practice survey,'' in \emph{2008 23rd IEEE/ACM International
  Conference on Automated Software Engineering - Workshops}, 2008, pp. 35--42.

\bibitem{Razavian2021SurveySOA}
M.~Razavian and P.~Lago, ``A survey of soa migration in industry,'' in
  \emph{Service-Oriented Computing}, G.~Kappel, Z.~Maamar, and H.~R.
  Motahari-Nezhad, Eds.\hskip 1em plus 0.5em minus 0.4em\relax Berlin,
  Heidelberg: Springer Berlin Heidelberg, 2011, pp. 618--626.

\bibitem{Flauzino:2018:YSS}
\BIBentryALTinterwordspacing
M.~Flauzino, J.~Ver\'{\i}ssimo, R.~Terra, E.~Cirilo, V.~H.~S. Durelli, and
  R.~S. Durelli, ``Are you still smelling it?: A comparative study between java
  and kotlin language,'' in \emph{Proceedings of the VII Brazilian Symposium on
  Software Components, Architectures, and Reuse}, ser. SBCARS '18.\hskip 1em
  plus 0.5em minus 0.4em\relax New York, NY, USA: ACM, 2018, pp. 23--32.
  [Online]. Available: \url{http://doi.acm.org/10.1145/3267183.3267186}
\BIBentrySTDinterwordspacing

\bibitem{Ardito2020KotlinJava}
\BIBentryALTinterwordspacing
L.~Ardito, R.~Coppola, G.~Malnati, and M.~Torchiano, ``Effectiveness of kotlin
  vs. java in android app development tasks,'' \emph{Information and Software
  Technology}, vol. 127, p. 106374, 2020. [Online]. Available:
  \url{http://www.sciencedirect.com/science/article/pii/S0950584920301439}
\BIBentrySTDinterwordspacing

\bibitem{Teyton:2014:SLM}
\BIBentryALTinterwordspacing
C.~Teyton, J.-R. Falleri, M.~Palyart, and X.~Blanc, ``A study of library
  migrations in java,'' \emph{J. Softw. Evol. Process}, vol.~26, no.~11, pp.
  1030--1052, Nov. 2014. [Online]. Available:
  \url{http://dx.doi.org/10.1002/smr.1660}
\BIBentrySTDinterwordspacing

\bibitem{Kula:2018:DUL}
\BIBentryALTinterwordspacing
R.~G. Kula, D.~M. German, A.~Ouni, T.~Ishio, and K.~Inoue, ``Do developers
  update their library dependencies?'' \emph{Empirical Softw. Engg.}, vol.~23,
  no.~1, pp. 384--417, Feb. 2018. [Online]. Available:
  \url{https://doi.org/10.1007/s10664-017-9521-5}
\BIBentrySTDinterwordspacing

\bibitem{hora2018sqj}
\BIBentryALTinterwordspacing
A.~Hora, R.~Robbes, M.~T. Valente, N.~Anquetil, A.~Etien, and S.~Ducasse, ``How
  do developers react to api evolution? a large-scale empirical study,''
  \emph{Software Quality Journal}, vol.~26, no.~1, pp. 161--191, Mar 2018.
  [Online]. Available: \url{https://doi.org/10.1007/s11219-016-9344-4}
\BIBentrySTDinterwordspacing

\bibitem{Teyton2013Discovering}
C.~{Teyton}, J.~{Falleri}, and X.~{Blanc}, ``Automatic discovery of function
  mappings between similar libraries,'' in \emph{2013 20th Working Conference
  on Reverse Engineering (WCRE)}, 2013, pp. 192--201.

\bibitem{Alrubaye:2019:UIR}
\BIBentryALTinterwordspacing
H.~Alrubaye, M.~W. Mkaouer, and A.~Ouni, ``On the use of information retrieval
  to automate the detection of third-party java library migration at the method
  level,'' in \emph{Proceedings of the 27th International Conference on Program
  Comprehension}, ser. ICPC '19.\hskip 1em plus 0.5em minus 0.4em\relax
  Piscataway, NJ, USA: IEEE Press, 2019, pp. 347--357. [Online]. Available:
  \url{https://doi.org/10.1109/ICPC.2019.00053}
\BIBentrySTDinterwordspacing

\bibitem{Salza:2018:DUT}
\BIBentryALTinterwordspacing
P.~Salza, F.~Palomba, D.~Di~Nucci, C.~D'Uva, A.~De~Lucia, and F.~Ferrucci, ``Do
  developers update third-party libraries in mobile apps?'' in
  \emph{Proceedings of the 26th Conference on Program Comprehension}, ser. ICPC
  '18.\hskip 1em plus 0.5em minus 0.4em\relax New York, NY, USA: ACM, 2018, pp.
  255--265. [Online]. Available:
  \url{http://doi.acm.org/10.1145/3196321.3196341}
\BIBentrySTDinterwordspacing

\end{thebibliography}

\bibliographystyle{IEEEtran} 

\vspace{-1.5cm}

\begin{IEEEbiography}[{\includegraphics[width=1in,height=1.25in,clip,keepaspectratio]{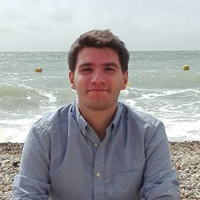}}]{Matias Martinez}
is an associate professor in the Universit\'e Polytechnique Hauts-de-France (France), and member of the LAMIH laboratory (UMR CNRS 8201). He got his PhD degree from University of Lille (France) and a Computer Science degree from UNICEN (Argentina). 

\end{IEEEbiography}
\vspace{-1.5cm}
\begin{IEEEbiography}[{\includegraphics[width=1in,height=1.25in,clip,keepaspectratio]{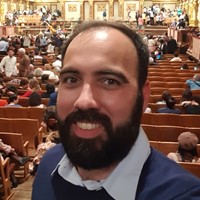}}]{
Bruno G\'ois Mateus} is an assistant professor of Software Engineering at the Federal University of Ceara - Campus Quixad\'a, Quixad\'a – CE, Brazil. 
He got his PhD from the Universit\'e Polytechnique Hauts-de-France, France.
His work focuses on empirical software engineering and studying quality of mobile applications.
\end{IEEEbiography}
\end{document}